\documentclass[12pt]{iopart}
\usepackage{graphicx}

\begin{document}

\title[]{Efficient production of polar molecular Bose-Einstein condensates
via an all-optical R-type atom-molecule adiabatic passage}

\author{Jing Qian$^{1,2}$, Lu Zhou$^{1}$, Keye Zhang$^{1}$ and Weiping Zhang$^{1,\ast}$}
\address{$^{1}$State Key Laboratory of Precision Spectroscopy, Department of Physics, East
China Normal University, Shanghai 200062, P. R. China}
\address{$^{2}$Department of Physics and Astronomy, Rowan University, Glassboro, New Jersey
08028, USA}
\ead{$^{\ast}$wpzhang@phy.ecnu.edu.cn}
\begin{abstract}
We propose a scheme of "$R$-type" photoassociative adiabatic passage (PAP)
to create polar molecular condensates from two different species of
ultracold atoms. Due to the presence of a quasi-coherent population trapping
state in the scheme, it is possible to associate atoms into molecules with a 
\textit{low-power} photoassociation (PA) laser. One remarkable advantage of
our scheme is that a tunable atom-molecule coupling strength can be achieved
by using a time-dependent PA field, which exhibits larger flexibility than
using a tunable magnetic field. In addition, our results show that the PA
intensity required in the "$R$-type" PAP could be greatly reduced compared
to that in a conventional "$\Lambda $-type" one.

\end{abstract}

\pacs{03.75.Mn,05.30.Jp,32.80.Qk}
\submitto{\NJP}
\maketitle

\section{Introduction}

Recently, the realization of ultracold polar molecular gases has been
regarded as one of the most promising research directions in the field of
atomic and molecular physics \cite{Doyle04,Summary}. Ultracold polar
molecules, with their long-range and anisotropic dipole-dipole interactions 
\cite{Santos00,Syi00}, have attracted much attention in a variety of
research areas, such as quantum information science \cite{DeMille02}-\cite%
{Yelin06} and precision measurement \cite{Kozlov95}-\cite{DeMille08}.

There are two typical routes to achieve quantum degenerate gases of
molecules. One is through the direct cooling of molecules, which is hard to
achieve due to the complex internal levels of molecules \cite{Doyle98}. The
alternative one is to couple a pair of degenerate atoms by photoassociation
(PA) \cite{Jones06} or Feshbach resonance (FR) \cite{Thorsten06}. However
the diatomic molecule formed by a PA or FR process is usually loosely bound
and energetically unstable. They have to be adiabatically transferred into a
tightly bound ground state via a stimulated Raman adiabatic passage (STIRAP) 
\cite{Bergmann98}. The success of the STIRAP is based on a coherent
population trapping (CPT) state, which is accomplished by a pair of pulses
in a counterintuitive sequence. During the adiabatic transfer, the system
can follow the superposition between initial and final states, preventing
any incoherent losses involving the middle unstable levels. Thereby, the
high phase-space density of the initial gas can be coherently preserved.

Currently, intensive experimental efforts to obtain quantum degenerate gases
of molecules have been made by combining FR with STIRAP, which serves as an
effective way to produce molecules in lower vibrational levels \cite%
{Winkler07}-\cite{Aikawa09}. However, due to the strong vibrational
quenching, a severe particle loss appears near the FR threshold. One way to
solve this problem is to apply the optical lattice technique \cite%
{Winkler07,Lang08}, in which inelastic collisions between molecules are well
suppressed by preparing one single molecule per lattice site. Alternatively,
all-optical transfer of molecules toward quantum degeneracy using a
"two-color PA" method has been demonstrated experimentally \cite{Wynar00}-%
\cite{Sage05}, where the excited molecules are moved down by a coherent dump
field, instead of by spontaneous decay \cite{Kerman04}-\cite{Deiglmayr08}.
However, one common bottleneck with PAs is the small free-bound
Franck-Condon factor (FCF), which requires an intense PA power to achieve an
efficient adiabatic transfer \cite{Vardi97,Vardi99}. To date, the most
promising way to overcome this PA weakness is by the FR-assisted PA scheme
proposed first by Verhaar \textit{et. al. }\cite{Abeelen98,Courteille98} and
verified by many groups later on \cite{Tolra03}-\cite{Kuznetsova09}. In
terms of these studies, if a Feshbach quasi-bound state is adjusted close to
the continuum, the atomic scattering wavefunction, acquiring some
bound-state properties, becomes more localized. This gives rise to a
dramatic enhancement of the free-bound FCF. As a result, the PA intensity
required for a given atom-molecule transfer efficiency can be greatly
reduced, compared with the case without the assistance of FR.

In the present work, we propose an all-optical scheme to achieve a high
transfer efficiency of atoms into molecules with a low PA power. For the
purpose, we consider a "$R$-type" atom-molecule conversion model (see figure %
\ref{model} in solid arrows) through a photoassociative STIRAP procedure.
Such a model is similar to the "$R$-transfer" suggested by Nikolov \textit{%
et. al. }\cite{Nikolov00} as well as to the work by Band and Julienne \cite%
{Band95}. In their works, molecules with an upper high-lying state are
generated first through a step-wise PA excitation from free atoms (e.g. $%
\left\vert 0_{1,2}\right\rangle \rightarrow \left\vert m\right\rangle
\rightarrow \left\vert e\right\rangle $ in figure \ref{model}), followed by
a radiative decay to populate a series of ground manifolds (e.g. $\left\vert
e\right\rangle \rightarrow \left\vert g\right\rangle $ in figure \ref{model}%
). In this paper, we apply a coherent dump field to make the transition from
state $\left\vert e\right\rangle $ to state $\left\vert g\right\rangle $,
instead of by spontaneous emission. This leads to an accessible
atom-molecule adiabatic passage between the initial ($\left\vert
0_{1,2}\right\rangle $) and target states ($\left\vert g\right\rangle $).
Compared with a conventional "$\Lambda $-type" model (see figure \ref{model},
dash-dotted arrows), the CPT state supported in the "$R$-type" scheme has
been perturbed by a newly embedded state $\left\vert m\right\rangle $, which
is absent in previous STIRAPs. As we will show, state $\left\vert
m\right\rangle $ can help to reduce the power in PA field, whose stability
properties will play a significant role in the molecular production. Under a
simple numerical comparison, we have identified that the PA power required
in the "$\Lambda $-type" model must be much higher than that in the "$R$%
-type" model for achieving the same final efficiency.

This paper is organized as follows. In section 2, after briefly reviewing a
similar idea of FR-induced STIRAP, we come up with our photoassociative
STIRAP model and develop the underlying mean-field equations for the studies
of a quasi-CPT description. In section 3, a generalized adiabatic theorem
involving all the Bogoliubov collective modes is introduced to evaluate the
adiabatic condition and quasi-CPT lifetime in our scheme. In section 4,
numerical simulations for the cases described in section 2 and 3 are
implemented by using practical parameters. The laser profiles applied in the
calculations are optimized according to adiabatic condition (section 4.1)
and other relevant assumptions (section 4.2). Finally, a summary is given in
section 5.

\section{Model and dark state theory}

\begin{figure}
[ptb]
\begin{center}
\includegraphics[
height=2.45in, width=2.5851in
]%
{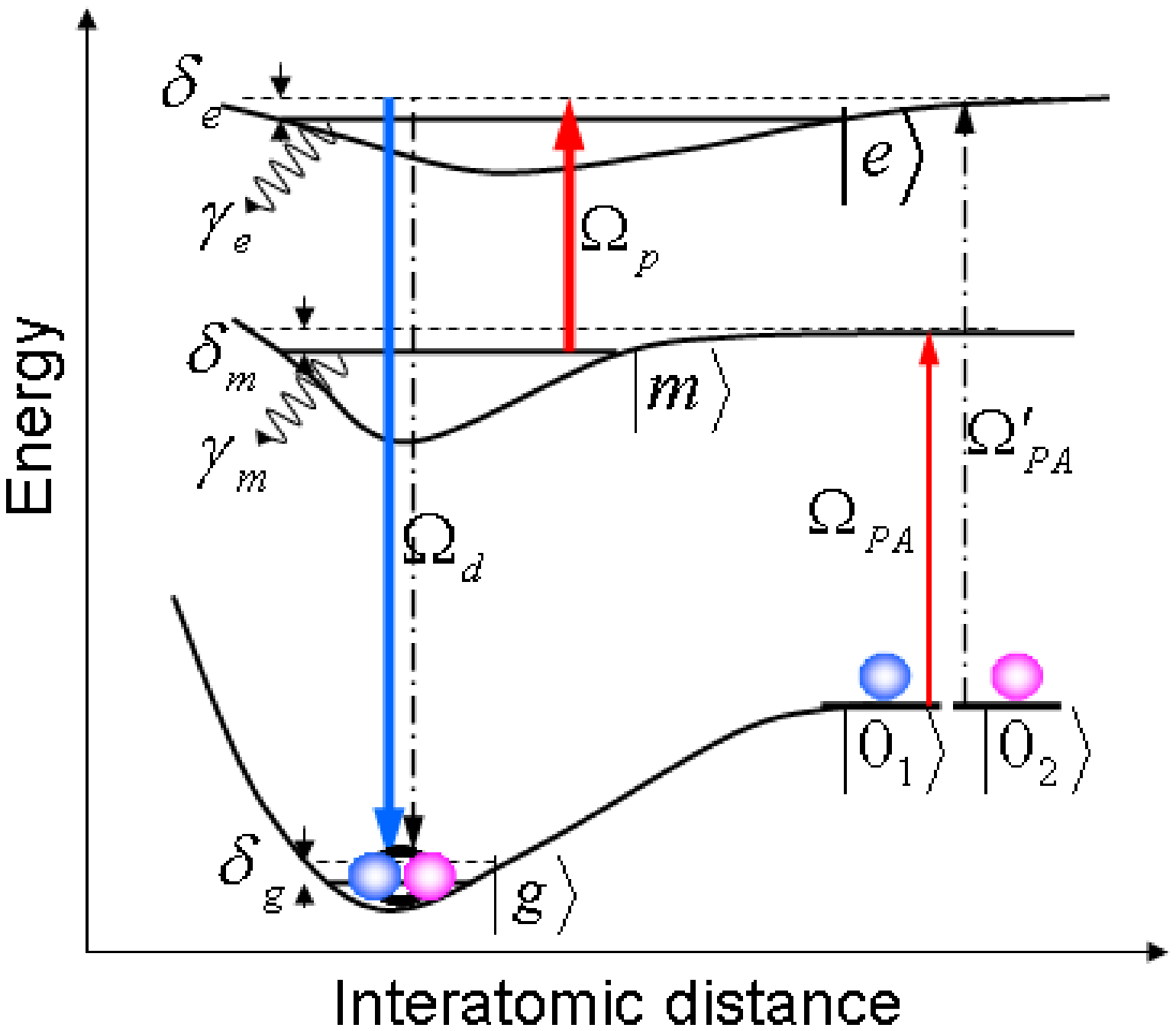}%
\caption{(color online) Schematic diagram: "$R$-type" transfer in
solid arrows comprising the transitons of $\left\vert
0_{1,2}\right\rangle \rightarrow \left\vert m\right\rangle
\rightarrow\left\vert e\right\rangle \rightarrow \left\vert
g\right\rangle $ with the corresponding coupling fields
$\Omega_{PA}$, $\Omega_{p}$ and $\Omega_{d}$, respectively;
"$\Lambda$-type" transfer in dash-dotted arrows with $\left\vert
0_{1,2}\right\rangle \rightarrow\left\vert e\right\rangle
\rightarrow\left\vert g\right\rangle $ transitions characterized by
$\Omega_{PA}^{\prime}$ and $\Omega_{d}$. All the
other parameters are described in the text.}%
\label{model}%
\end{center}
\end{figure}

Before moving to concrete illustrations of the model, we briefly review the
idea of the FR-induced STIRAP method \cite{Kokkelmans01,Mackie02}. In a
typical FR-induced STIRAP, a number of colliding atoms undergo a strong
association into quasi-bound molecules when a magnetic field is swept close
to or across the FR, characterized by coupling strength $\alpha $ and
binding energy $\varepsilon $. Subsequently, these quasi-bound molecules are
further transferred into stable molecules via a STIRAP. Clearly, here $%
\alpha $ and $\varepsilon $ correspond to $\Omega _{PA}$ and $\delta _{m}$
(see figure \ref{model}) in our scheme, respectively. A major advantage of
our scheme is that $\Omega _{PA}$ and $\delta _{m}$ can be manipulated more
conveniently than $\alpha $ and $\varepsilon $ over the time scales. Because
the latter quantities are highly dependent on the atomic intrinsic
properties, especially the coupling strength $\alpha $, which is fixed by the
hyperfine interaction and is hence independent of time, $\varepsilon $ is
experimentally tunable via an external magnetic field, while the former
quantities, being controllable by optical means, are easily selected.

Turning to our scheme, as depicted in figure \ref{model}, we study a "$R$%
-type" five-level atom-polar-molecule formation. Two species of free atoms
prepared in $\left\vert 0_{1}\right\rangle $ and $\left\vert
0_{2}\right\rangle $ states are first coupled into molecules in an
intermediate high-lying state $\left\vert m\right\rangle $ with Rabi
frequency $\Omega _{PA}$ and detuning $\delta_{m}$. Simultaneously, a pair
of pump-dump lasers are applied to move these loosely bound molecules in $%
\left\vert m\right\rangle $ down to the lowest molecular ground state $%
\left\vert g\right\rangle $, where $\Omega_{p}$, $\Omega_{d}$ stand for
coupling strengths and $\delta_{e}$, $\delta_{g}$ for the two- and
three-photon detunings. This scheme has several attractive properties.
Firstly, the presence of state $\left\vert m\right\rangle $ brings one extra
bound-bound transition from state $\left\vert m\right\rangle $ to state $%
\left\vert e\right\rangle $; hence it becomes easier for the PA field to
associate atoms into the $\left\vert m\right\rangle $, rather than a higher $%
\left\vert e\right\rangle $ state. Secondly, transitions $\left\vert
m\right\rangle \rightarrow\left\vert e\right\rangle $ and $\left\vert
e\right\rangle \rightarrow\left\vert g\right\rangle $ are preferred because
of the favorable bound-bound FCFs. Meanwhile, the $\left\vert
0_{1,2}\right\rangle \rightarrow\left\vert m\right\rangle $ transition is
also accessible by an optimal control of the PA field in the time domain.

As usual, we start our discussions with a set of coupled Gross-Pitaevskii's
equations. In the mean-field treatment, where every quantum field operator $%
\hat{\Psi}_{i}$ has been replaced by its normalized amplitude $\psi_{i}$ 
\cite{Heinzen00}, this yields%

\begin{eqnarray}
i\dot{\psi}_{0_{1}}  &
=-\frac{\Omega_{PA}}{2}\psi_{0_{2}}^{\ast}\psi
_{m},\label{dy1}\\
i\dot{\psi}_{0_{2}}  &
=-\frac{\Omega_{PA}}{2}\psi_{0_{1}}^{\ast}\psi
_{m},\label{dy2}\\
i\dot{\psi}_{m}  & =-\left(  \delta_{m}+i\gamma_{m}\right)  \psi_{m}%
-\frac{\Omega_{PA}}{2}\psi_{0_{1}}\psi_{0_{2}}-\frac{\Omega_{p}}{2}\psi
_{e},\label{dy3}\\
i\dot{\psi}_{e}  & =-\left(  \delta_{e}+i\gamma_{e}\right)  \psi_{e}%
-\frac{\Omega_{p}}{2}\psi_{m}-\frac{\Omega_{d}}{2}\psi_{g},\label{dy4}\\
i\dot{\psi}_{g}  & =-\delta_{g}\psi_{g}-\frac{\Omega_{d}}{2}\psi
_{e},\label{dy5}%
\end{eqnarray}
where $\gamma_{i}$ (i=m,g) is introduced phenomenologically to
describe the spontaneous decay of the $\left\vert i\right\rangle $ state
to other undetected states, and it is possible to find a relatively
stable $\left\vert m\right\rangle $ state with its decay rate
$\gamma_{m}\ll\gamma_{e}$ \cite{Band95,Napolitano94}. The initial
and target states are assumed to be sufficiently stable with
$\gamma_{0_{1(2)},g}\equiv0$. For an easy analysis without loss of
the main physics, inter- and intra-species collisions have been ignored
under typical parameters \cite{Kuznetsova08}. After a global gauge
transformation, we can safely consider all the Rabi frequencies to
be real positive values without loss of generality.

A CPT state is always expected to move all the population into a
target state as long as the adiabatic condition holds. In order to
derive the corresponding adiabatic parameter, we first search for
the CPT distributions for the
following assumptions:%

\begin{equation}
\psi_{0_{1,2}}=\phi_{0}e^{-i\mu
t},\psi_{e}=0,\psi_{m,g}=\phi_{m,g}e^{-2i\mu
t}. \label{ansatz}%
\end{equation}

Here $\phi_{i}$ is a steady-state amplitude, we consider $\phi_{0_{1}}%
=\phi_{0_{2}}=\phi_{0}$ for a balanced system, and $\mu$ is the
atomic chemical potential. By ignoring all the decays and inserting
equation (\ref{ansatz}) into equations (\ref{dy1})-(\ref{dy5}) with
particle number conservation:
$2(\phi_{0}^{2}+\phi_{m}^{2}+\phi_{g}^{2})=1$, a generalized
three-photon
resonance is given by%

\begin{equation}
\delta_{g\pm}=-2\mu_{\pm}=\frac{-\Omega_{PA}^{2}/2}{\delta_{m}\pm\sqrt
{\delta_{m}^{2}+\Omega_{PA}^{2}\left(  3+\chi^{2}\right)  /2}},
\label{delta_g}%
\end{equation}
leading to the following CPT descriptions with $\phi_{e}=0$:%

\begin{eqnarray}
\phi_{0}  &  =\sqrt{\frac{1}{2}-\phi_{m}^{2}\left(  1+\chi^{2}\right)  },\label{CPT1}\\
\phi_{m}  &  =-\frac{\bar{\Omega}_{PA}/2}{1+\sqrt{1+\bar{\Omega}_{PA}%
^{2}\left(  3+\chi^{2}\right)  /2}},\label{CPT2}\\
\phi_{g}  &  =-\chi\phi_{m}. \label{CPT3}%
\end{eqnarray}
where $\bar{\Omega}_{PA}=\Omega_{PA}/\delta_{m}$, $\chi=\Omega_{p}/\Omega_{d}%
$. In equation (\ref{delta_g}), the choice of $\delta_{g}$ is
determined by $\delta_{m}$. If $\delta_{m}>0$,
$\delta_{g}=\delta_{g+}$ and $\mu=\mu_{+}$, whereas if $\delta_{m}<0$,
$\delta_{g}=\delta_{g-}$ and $\mu=\mu_{-}$. From equations
(\ref{CPT1})-(\ref{CPT3}), we note that when $\left\vert
\bar{\Omega }_{PA}\right\vert $ and $\chi$ both change from 0 to
large positive values, population initially prepared in states
$\left\vert 0_{1,2}\right\rangle $ will be gradually converted into
molecules in state $\left\vert g\right\rangle $ under three-photon
resonance [equation (\ref{delta_g})]. Also it is worth emphasizing
that such a CPT state has been perturbed since
$\phi_{m}\neq0$, and is called a "quasi-CPT" state. In the limit
of $\bar{\Omega}_{PA}\ll1$, population in state $\left\vert
m\right\rangle $ is virtually empty, we find that a complete
transfer is still possible as long as $\chi$ varies from 0 to
$\infty$. In other words, the change by $\bar{\Omega}_{PA}$ has been
accomplished by varying $\Omega_{p}$; thus the existence of state
$\left\vert m\right\rangle $ is\ quite helpful for a relatively
small $\bar{\Omega}_{PA}$ value.

Actually, in the dynamics, if we use a strong PA laser to trigger
the $\left\vert 0_{1,2}\right\rangle \rightarrow\left\vert
m\right\rangle $ transition, particle accumulations in state
$\left\vert m\right\rangle $ will inevitably arise. Therefore,
in order to avoid a considerable loss from state $\left\vert
m\right\rangle $, pulse durations in STIRAP must be much shorter
than $\left\vert m\right\rangle $ state's lifetime. On the other
hand, if we deeply reduce the $\Omega_{PA}$ value, the population in the
$\left\vert m\right\rangle $ state will greatly be suppressed;
meanwhile, a large fraction of atoms are left in the continuum,
unpaired, because of a poor atom-molecule coupling strength. This
conflict can be generalized to the properties of a quasi-CPT state,
in which case one may prefer the use of moderate PA power.

Results in equations (\ref{CPT1})-(\ref{CPT3}) are for the case of
$\delta_{m}\neq0$. If $\delta_{m}=0$, i.e. the PA laser is exactly
resonant with the free-bound transition, then equation
(\ref{delta_g}) is reduced to $\delta_{g\pm}=\pm
\Omega_{PA}/(6+2\chi^{2})^{1/2}$\ with the following CPT solutions:%

\begin{equation}
\phi_{0}^{2}=2\phi_{m}^{2}=2\phi_{g}^{2}/\chi^{2}=\left(
3+\chi^{2}\right)
^{-1}. \label{CPT_re}%
\end{equation}

Equation (\ref{CPT_re}) shows a constant population ratio
between states $\left\vert 0_{1,2}\right\rangle $ and $\left\vert
m\right\rangle $, i.e. $\phi_{0}^{2}/\phi_{m}^{2}=2$. This equality
contrasts with the standard CPT evolution, especially at t=0,
which implies a poor transfer efficiency at $\delta_{m}=0$. As a
result, a nonzero $\delta_{m}$ value is favored in our
consideration.

\section{Adiabatic Theorem}

To derive the adiabatic parameter for the quasi-CPT state, we adopt
a standard linearized approach as in \cite{Pu07,Jing08} by
adding a small fluctuation $\delta\psi_{i}$ to the
instantaneous steady-state solution $\phi_{i}$,%
\begin{equation}
\psi _{0_{1,2}}=\left( \phi _{0}+\delta \psi _{0_{1,2}}\right)
q\left( t\right) ,\psi _{e}=\delta \psi _{e}q^{2}\left( t\right)
,\psi _{m,g}=\left(
\phi _{m,g}+\delta \psi _{m,g}\right) q^{2}\left( t\right) \label{linearizeda}%
\end{equation}
where $q\left(  t\right)  =\exp[-\int_{0}^{t}\mu\left(
t^{\prime}\right) dt^{\prime}]$, and $\mu(t)$ is a time-dependent
chemical potential given by $\mu\left(  t\right)  =\mu_{+\left(
-\right)  }$ (see equation (\ref{delta_g})). Substituting equation
(\ref{linearizeda}) into the mean-field equations
(\ref{dy1})-(\ref{dy5}) with the help of CPT descriptions, we
eventually arrive at a set of linearized equations for the vector
$\mathbf{\delta}\mathbf{\psi=}\left[  \delta
\psi_{0_{1}},\delta\psi_{0_{2}},\delta\psi_{m},\delta\psi_{e},\delta\psi
_{g}\right]  ^{T}$ with its conjugate vector
$\mathbf{\delta}\mathbf{\psi
}^{\ast}$%

\begin{equation}
\mathbf{\delta\dot{\Psi}=-}i\mathbf{M\delta\Psi-\Gamma\delta\Psi-\dot{\Phi}},
\label{linearized}%
\end{equation}
where
\[
\mathbf{M=}\left(
\begin{array}
[c]{cc}%
\mathbf{A} & \mathbf{B}\\
-\mathbf{B} & -\mathbf{A}%
\end{array}
\right)  ,\mathbf{\Gamma=}\left(
\begin{array}
[c]{cc}%
\mathbf{\gamma} & 0\\
0 & \mathbf{\gamma}%
\end{array}
\right)  ,
\]
and
\begin{eqnarray}
\mathbf{A}  &  =-\frac{1}{2}\left(
\begin{array}
[c]{ccccc}%
2\mu & 0 & \Omega_{PA}\phi_{0} & 0 & 0\\
0 & 2\mu & \Omega_{PA}\phi_{0} & 0 & 0\\
\Omega_{PA}\phi_{0} & \Omega_{PA}\phi_{0} & 2\left(
\delta_{m}+2\mu\right)  &
\Omega_{p} & 0\\
0 & 0 & \Omega_{p} & 0 & \Omega_{d}\\
0 & 0 & 0 & \Omega_{d} & 0
\end{array}
\right)  ,\\
\mathbf{B}  &  =-\frac{\Omega_{PA}\phi_{m}}{2}\left(
\begin{array}
[c]{ccccc}%
0 & 1 & 0 & 0 & 0\\
1 & 0 & 0 & 0 & 0\\
0 & 0 & 0 & 0 & 0\\
0 & 0 & 0 & 0 & 0\\
0 & 0 & 0 & 0 & 0
\end{array}
\right).
\end{eqnarray}

In equation (\ref{linearized}), some notations are
$\mathbf{\delta\Psi=}\left[
\mathbf{\delta}\mathbf{\psi,\delta}\mathbf{\psi}^{\ast}\right]
^{T}$, $\mathbf{\dot{\Phi}=}\left[
\mathbf{\dot{\phi},\dot{\phi}}\right]  ^{T}$ with
$\mathbf{\dot{\phi}=}\left[
\dot{\phi}_{0},\dot{\phi}_{0},\gamma_{m}\phi
_{m}+\dot{\phi}_{m},0,\dot{\phi}_{g}\right]  ^{T}$.
$\mathbf{\gamma}$ is a $5\times5$ matrix with
$\gamma_{33}=\gamma_{m}$ and $\gamma_{44}=\gamma_{e}$ being the only
nonzero elements. In addition, we have assumed detunings
$\delta_{e,g}=-2\mu(t)$ to be chirped \cite{Koenig04}.

Furthermore, we introduce a generalized Bogoliubov-de Gennes (BdG)
equation for
matrix $\mathbf{M}$,%

\begin{equation}
\mathbf{M}\left(  t\right)  \mathbf{w}_{i}\left(  t\right)
=\omega_{i}\left(
t\right)  \mathbf{w}_{i}\left(  t\right)  , \label{BdG}%
\end{equation}
where $\omega_{i}$ and $\mathbf{w}_{i}=\left[  \mathbf{u}_{i},\mathbf{v}%
_{i}\right]  ^{T}$ are the well-defined $i$th eigenenergy and
eigenvector,
respectively. $\mathbf{u}_{i}$ and $\mathbf{v}_{i}$ contain familiar
Bogoliubov
$u$-$v$ parameters for each species%

\begin{equation}
\mathbf{u}\left(  \mathbf{v}\right)  _{i}=[u\left(  v\right)  _{i0_{1}%
},u\left(  v\right)  _{i0_{2}},u\left(  v\right)  _{im},u\left(
v\right)
_{ie},u\left(  v\right)  _{ig}]^{T}. \label{uv}%
\end{equation}

From the BdG equation, taking into account the special structure of matrix \textbf{M},
one can show the quantities $\omega_{i}^{2}$ are the eigenenergies of the matrix $(\mathbf{A}%
+\mathbf{B})(\mathbf{A}-\mathbf{B})$, which can be obtained from the
following
equation:%
\begin{equation}
\left(  \omega_{i}^{2}\right)  ^{2}\left(  \left(
\omega_{i}^{2}\right)
^{3}-a_{1}\left(  \omega_{i}^{2}\right)  ^{2}+a_{2}\left(  \omega_{i}%
^{2}\right)  -a_{3}\right)  =0, \label{cubic}%
\end{equation}
where the coefficients $a_{i}$ are given as%

\begin{eqnarray}
a_{1}  &  =\left(  \delta_{m}+2\mu\right)  \left(
\delta_{m}+6\mu\right)
+\frac{\Omega_{p}^{2}+\Omega_{d}^{2}}{2},\\
a_{2}  &  =\frac{\left(  \delta_{m}+2\mu\right)  }{2}\left(  2\Omega_{p}%
^{2}\mu+\Omega_{d}^{2}\left(  \delta_{m}+6\mu\right)  \right)
+\frac{\left(
\Omega_{p}^{2}+\Omega_{d}^{2}\right)  ^{2}}{16},\\
a_{3}  &  =\frac{\left(  \delta_{m}+2\mu\right)
\Omega_{d}^{2}}{16}\left( 4\Omega_{p}^{2}\mu+\Omega_{d}^{2}\left(
\delta_{m}+6\mu\right)  \right)  .
\end{eqnarray}

Eigenenergies implied in equation (\ref{cubic}) comprise a doublet
0 mode $\omega_{0,1}=0$ and three pairs of excited modes $\left(
\omega _{j},-\omega_{j}^{\ast}\right)  $ (j=2,3,4). We find that
$\omega_{j}$ is real and has to be determined by biorthonormal
relations for its corresponding eigenvector $\mathbf{w}_{j}$.
Detailed elucidations of biorthonormality have been published elsewhere
\cite{Ling07}. In addition, we realize that the dynamical
instability is impossible here due to the absence of collisions.

To accomplish the goal of deriving the adiabatic theorem, we have to
expand an arbitrary vector $\mathbf{\delta\Psi}$ in the
dressed-state picture with a complete set of eigenvectors. By
solving the BdG equation with 0 eigenenergies, we are able to
obtain $\mathbf{w}_{0}$ and $\mathbf{w}_{1}$ (unnormalized) explicitly using
Gram-Schmidt orthogonalization,

\begin{eqnarray}
\mathbf{w}_{0}  &  =\left(  -1,1,0,0,0,1,-1,0,0,0\right)  ^{T},\label{w0}\\
\mathbf{w}_{1}  &  =\left(  \frac{\phi_{0}}{2\phi_{m}},\frac{\phi_{0}}%
{2\phi_{m}},1,0,-\chi,\frac{-\phi_{0}}{2\phi_{m}},\frac{-\phi_{0}}{2\phi_{m}%
},-1,0,\chi\right)  ^{T}. \label{w1}%
\end{eqnarray}
Here $\mathbf{w}_{0}$, being a real dark state, is entirely
decoupled with other eigenmodes, because its source term
($\mathbf{w}_{0}^{T}\mathbf{\dot {\Phi}}$) and inter-coupling term
($\mathbf{w}_{0}^{T}\mathbf{\Gamma \mathbf{\eta}_{+}w}_{j}$) both
vanish (see equation (\ref{cj}) below for detailed notations),
whereas, $\mathbf{w}_{1}$ is most likely to be triggered through its
nonzero inter-coupling strength ($\mathbf{w}_{1}^{T}\mathbf{\Gamma
\mathbf{\eta}_{+}w}_{j}$), which can be roughly estimated by its decay rate%
\begin{equation}
\mathbf{w}_{1}^{T}\mathbf{\Gamma\eta}_{+}\mathbf{Q}=\frac{\gamma_{m}}{\phi
_{0}^{2}/2\phi_{m}^{2}+2(1+\chi^{2})}\equiv1/\tau_{cpt}, \label{cpt_lifetime}%
\end{equation}
where $\mathbf{Q}$ is a newly introduced vector complementary to
$\mathbf{w}_{1}$ with a well-defined normalization,%

\begin{equation}
\mathbf{w}_{1}^{T}\mathbf{\eta}_{+}\mathbf{Q}=1,\label{normal}%
\end{equation}
through the definition of%

\begin{equation}
\mathbf{MQ=w}_{1}/v.\label{v_d}%
\end{equation}

Here, $v$ is a coefficient to be determined, and $\mathbf{\eta}_{+}$(and
$\mathbf{\eta}_{-}$ below) are given in \cite{Ling07}.
Combining equation (\ref{normal}) with (\ref{v_d}), we find the
vector $\mathbf{Q}$ takes a
special form: $\mathbf{Q}=\mathbf{[}q_{0},q_{0},q_{m},q_{e},q_{g},q_{0}%
,q_{0},q_{m},q_{e},q_{g}\mathbf{]}^{T}$. Detailed expressions for
$q_{i}$ and $v$ are presented in the appendix.

The CPT lifetime $\tau_{cpt}$ defined in equation
(\ref{cpt_lifetime}) is clearly inversely-proportional to
$\gamma_{m}$ and $\phi_{m}$, which agrees with our intuitions. In
other words, the presence of state $\left\vert m\right\rangle $
actually gives rise to a finite lifetime for the quasi-CPT state.
Any pulse duration used in the system has to be much shorter than
$\tau_{cpt}$; otherwise, a big particle loss from state $\left\vert
m\right\rangle $ is unavoidable. One effective way to achieve a long
$\tau_{cpt}$ is to search for a relatively stable $\left\vert
m\right\rangle $ state with a small $\gamma_{m}$ value. Other
excited eigenenergies and eigenvectors are too complicated to
list here, but they can be conveniently derived from equation
(\ref{BdG}) with (\ref{cubic}).

Since other inter-coupling strengths for $\mathbf{w}_{1}$ are also
proportional to $\gamma_{m}$ as in equation (\ref{cpt_lifetime}) and
$\gamma_{m}$ is considered to be much smaller than $\gamma_{e}$, we
shall safely ignore the contributions from $\mathbf{w}_{1}$ and
expand $\mathbf{\delta\Psi}$ in the parameter space with the help of
three excited eigenmodes $\mathbf{w}_{j}$
(j=2,3,4) only, taking the form of%

\begin{equation}
\mathbf{\delta\Psi=}\sum_{j=2}^{4}\left(  c_{j}\mathbf{\eta}_{+}\mathbf{w}%
_{j}-c_{j}^{\ast}\mathbf{\eta}_{-}\mathbf{w}_{j}^{\ast}\right).  \label{DPHI}%
\end{equation}

Through inserting equation (\ref{DPHI}) into (\ref{linearized}), and
with the help of biorthonormality relations, finally, we obtain a
set of coupling
equations for $c_{j}\left(  t\right)  $,%

\begin{equation}
\dot{c}_{j}+i\omega_{j}c_{j}+\mathbf{\mathbf{w}}_{j}^{T}\mathbf{\Gamma
\delta\Psi}=\mathbf{-\mathbf{w}}_{j}^{T}\mathbf{\dot{\Phi}} \label{cj}%
\end{equation}

Terms like $\mathbf{\dot{w}}_{j}^{T}\mathbf{\delta\Psi}$ have been
eliminated in equation (\ref{cj}) because the eigenvector
$\mathbf{\mathbf{w}}_{j}$ changes very slowly in the adiabatic
limit. Generally speaking, if a system is said to operate in an
adiabatic regime, population in any excited mode (nonzero
eigen-mode) remains small. Hence, we shall introduce a typical
adiabatic parameter definition
\begin{equation}
r\left(  t\right)  =\sqrt{\frac{\left\vert c_{2}\right\vert
^{2}+\left\vert
c_{3}\right\vert ^{2}+\left\vert c_{4}\right\vert ^{2}}{3}}\ll1\label{r_value}%
\end{equation}

A reduction in the $r$-value means an increase in the adiabaticity;
in general, it can be accomplished by a longer pulse or a stronger
pump field. In the adiabatic regime, if a system evolves in a
CPT state, an entire population conversion is achievable. However,
the CPT lifetime implied in our model places a limitation for both
the pulse duration and PA intensity, leading to a slightly larger
$r$-value. This point will be discussed in section 4.1 toward the
goal of obtaining an optimal pulse duration and PA intensity for an
efficient transfer.

In equation (\ref{cj}), since $\dot{c}_{j}$ can be ignored
adiabatically, we further rewrite it as a series of linearized
coupling
equations:%
\begin{equation}
\left(
\begin{array}
[c]{cc}%
\mathbf{F}+i\mathbf{\omega} & \mathbf{G}\\
\mathbf{G} & \mathbf{F}-i\mathbf{\omega}%
\end{array}
\right)  \left(
\begin{array}
[c]{c}%
\mathbf{c}\\
\mathbf{c}^{\ast}%
\end{array}
\right)  =-\left(
\begin{array}
[c]{c}%
\dot{\Phi}_{w}\\
\dot{\Phi}_{w}%
\end{array}
\right)  \label{coupled}%
\end{equation}
where%
\[
\mathbf{F=}\left(
\begin{array}
[c]{ccc}%
f_{22} & f_{23} & f_{24}\\
f_{23} & f_{33} & f_{34}\\
f_{24} & f_{34} & f_{44}%
\end{array}
\right)  ,\mathbf{G=}\left(
\begin{array}
[c]{ccc}%
0 & g_{23} & g_{24}\\
-g_{23} & 0 & g_{34}\\
-g_{24} & -g_{34} & 0
\end{array}
\right)  ,
\]
with the definitions of $\mathbf{c=}\left[  c_{2},c_{3},c_{4}\right]
^{T}$,
$\dot{\Phi}_{w}=[\mathbf{\mathbf{w}}_{2}^{T}\mathbf{\dot{\Phi},\mathbf{w}}%
_{3}^{T}\mathbf{\dot{\Phi},\mathbf{w}}_{4}^{T}\mathbf{\dot{\Phi}]}^{T}$,
$\mathbf{\omega}=\omega_{j}\mathbf{D}$ (j=2,3,4), $\mathbf{D}$ is a
$3\times3$ unit matrix, and

\begin{eqnarray}
f_{ij} &  =\gamma_{m}\left(  u_{im}u_{jm}-v_{im}v_{jm}\right)
+\gamma
_{e}\left(  u_{ie}u_{je}-v_{ie}v_{je}\right)  ,\label{f}\\
g_{ij} &  =\gamma_{m}\left(  v_{im}u_{jm}-u_{im}v_{jm}\right)
+\gamma
_{e}\left(  v_{ie}u_{je}-u_{ie}v_{je}\right)  ,\label{g}%
\end{eqnarray}

We solve $c_{j}(c_{j}^{\ast})$ values from equations (\ref{coupled})
numerically and insert them into equation (\ref{r_value}), a
time-dependent $r$-function is ultimately accessible. It needs to
be noted that all the $u(v)$s in equation
(\ref{f}) and (\ref{g}) have been normalized according to biorthogonality,%

\begin{equation}
\mathbf{w}_{i}^{T}\mathbf{\eta}_{+}\mathbf{w}_{j}=\delta_{ij},\mathbf{w}%
_{i}^{T}\mathbf{\eta}_{-}\mathbf{w}_{j}=0. \label{bio}%
\end{equation}

\section{Numerical Analysis}

In the following numerical calculations, we intended to achieve
a highly-efficient ground molecular production under an optimization
of all the optical fields, including $\Omega_{PA},$ $\Omega_{p}$ and
$\Omega_{d}$. From CPT descriptions [equations
(\ref{CPT1}-\ref{CPT3})], we adopt a common pair of
counterintuitive pump-dump pulses for $\left\vert m\right\rangle
$-$\left\vert
g\right\rangle $ transition with the same width $T$%

\begin{equation}
\Omega_{p,d}=\frac{\Omega_{p,d}^{0}}{2}\left(  1\pm\tanh\left(
\frac
{t-t_{p,d}}{T}\right)  \right)  \label{Rabi}%
\end{equation}
where $\Omega_{p,d}^{0}$, $t_{p,d}$ are for the peak Rabi
frequencies and central positions respectively. Based on equations
(\ref{CPT1})-(\ref{CPT3}), the PA Rabi frequency $\Omega_{PA}$,
which must start from 0, is considered to share the same profile as
$\Omega_{p}$ except for a different peak amplitude
$\Omega_{PA}^{0}$. Here, the detuning $\delta_{m}$ is fixed at a
finite value for simplicity.

\subsection{Optimal PA pulse}

In what follows, we seek to gain from the $r$-value in equation
(\ref{r_value}) insights into the parameters, especially for an
appropriate PA amplitude $\Omega_{PA}^{0}$ and a pulse duration $T$.
As we already understand, applying a longer pulse or a more intense PA
laser will lead to a lower $r$-value. If a system's adiabaticity
($r$-value) is kept in a low level, which means the system will
operate within the adiabatic regime, any excited modes are
greatly suppressed. In a pure-CPT environment, adiabaticity indeed
becomes a sufficient criterion for a complete transfer. However, in
our scheme, we observe in addition to adiabaticity, a long CPT lifetime
is another significant criterion for an efficient transfer.

\begin{figure}
[ptb]
\begin{center}
\includegraphics[
height=3.5434in, width=2.8532in
]%
{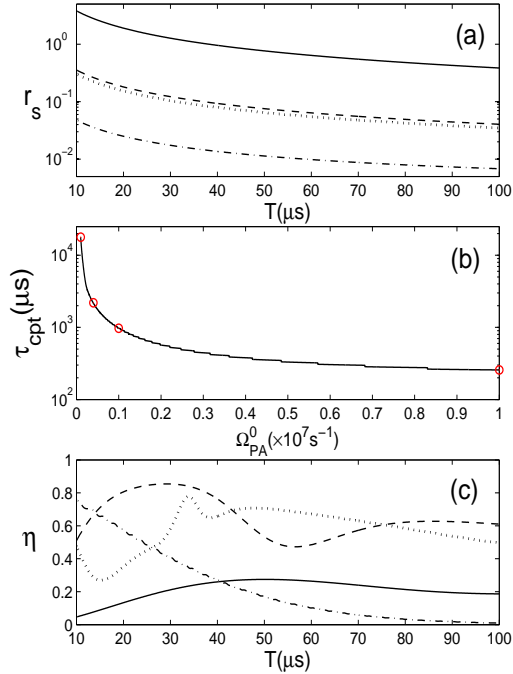}%
\caption{(a) $r$-value at $t=t_{s}$ versus pulse duration $T$ under
different PA amplitudes: from top to bottom $\Omega_{PA}^{0}=10^{5}$
s$^{-1}$, $4\times10^{5}$ s$^{-1}$, $10^{6}$ s$^{-1}$ and $10^{7}$
s$^{-1}$, respectively. (b) $\tau_{cpt}$ estimated at $t=t_{s}$ as a
function of $\Omega_{PA}^{0}$. The four circles (from left to right)
denote $\tau_{cpt}$=17.8 ms, 2.2 ms, 0.97 ms, 257 $\mu s$ with
respect to the corresponding $\Omega_{PA}^{0}$s shown in (a). (c)
Final molecular production $\eta =2|\psi_{g}(\infty)|^{2}$ versus pulse
duration $T$ under the same four cases as
in (a). The other parameters are described in the text.}%
\label{r_dy}%
\end{center}
\end{figure}

In a dynamical process, the $r$-value obtained from equation
(\ref{r_value}) varies with time. We find that $r$-value estimated
at $t_{s}$ which is defined by
$\phi_{0}^{2}(t_{s})=2\phi_{g}^{2}(t_{s})$ turns out to be a good
estimate of the degree of adiabaticity. Thus,
$r_{s}$ and $\tau_{cpt}$ values displayed in figure \ref{r_dy}(a), (b) are both evaluated at $t=t_{s}$.%

Figure \ref{r_dy}(a) and (c) present the variations of adiabaticity
$r(t_{s})$ and final efficiency $\eta$(=$\left\vert
2\psi_{g}(\infty)\right\vert ^{2}$) as a function of pulse width $T$,
respectively. As plotted in figure 2(a), either a longer pulse (from left to
right) or a stronger PA amplitude (from the top to the bottom) leads to an
improved adiabaticity. Furthermore, when $\Omega_{PA}^{0}$ is very
weak, such as 10$^{5}$ s$^{-1}$ (in solid), the $r$-value is around 1.0,
which cannot well satisfy the adiabatic condition $r\ll1$. Although
at this time, the CPT lifetime in figure \ref{r_dy}(b) is long
enough (more than 15 ms) to support a longer pulse duration, a large
part of atoms will be left in the continuum, resulting in poor
molecular production, which is no more than 30\% (see the solid curve in
(c)). On the other hand, if we use an intense laser,
$\Omega_{PA}^{0}=10^{7}$ s$^{-1}$, then the adiabaticity reduces
into the 0.01 level, whereas simultaneously, the CPT lifetime is only
around 250$\mu s$, leading to a dramatic reduction in $\eta$ as $T$
increases (see the dash-dotted curve in figure 2(c)), because with a longer $T$
value, a number of molecules decay spontaneously due to
$\gamma_{m}$. Obviously, if $T<20\mu s$, a relatively
higher $\eta$ value (%
$>$%
50\%) is still attainable.

In addition, we study two moderate cases with the PA amplitudes:
$\Omega _{PA}^{0}=4\times10^{5}$ s$^{-1}$ (in dashed) and
$\Omega_{PA}^{0}=10^{6}$ s$^{-1}$ (in dotted). No impressive
differences are observable in adiabaticity according to figure
\ref{r_dy}(a), where both are around the 0.1 level. Meanwhile,
the $\tau_{cpt}$ values represented in figure \ref{r_dy}(b) are both close to
1 ms, which do offer more space for a tunable $T$ value. Final
efficiencies in figure \ref{r_dy}(c) clearly exhibit a $T$-dependent
feature, while staying at a highly efficient level compared with two
former cases.

In light of the above discussions, we conclude that the adiabaticity
indeed serves as a useful tool to select favorable parameters.
Meanwhile, it is equivalently important to take the CPT
lifetime into
consideration. Here, we prefer to use $T=30$ $\mu s$, $\Omega_{PA}^{0}%
=4\times10^{5}$ s$^{-1}$.

\subsection{Optimal pump-dump pulse sequence}

The goal of this subsection is to design the optimal pump-dump
two-pulse sequence to maximize the yield of molecules.
Clearly, there are five individual parameters to be determined: $t_{p}$, $t_{d}$, 
$\Omega _{p}^{0}$, $\Omega _{d}^{0}$ and $\delta_{m}$. Such a
five-parameter variation is difficult to carry out. However, from
the CPT descriptions, we guess that the population
dynamics are most likely to be affected by the ratio $\chi=\Omega_{p}%
/\Omega_{d}$ instead of the $\Omega_{p}$ and $\Omega_{d}$ values. Therefore, we
introduce two new variables, which are $dt=t_{d}-t_{p}$ for pulse
delay and $\chi^{0}=\Omega _{p}^{0}/\Omega_{d}^{0}$ for peak
amplitude ratio. In combination with the one-photon detuning
$\delta_{m}$, there are three effective
quantities to be optimized.%

\begin{figure}
[ptb]
\begin{center}
\includegraphics[
height=3.2127in, width=2.5087in
]%
{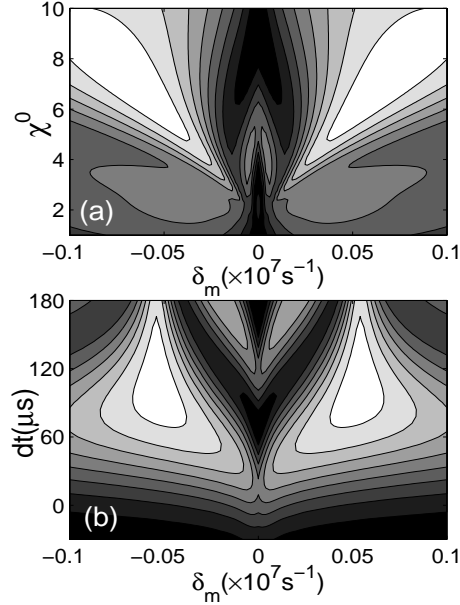}%
\caption{Contour plot of final transfer efficiencies under different
sets of (a) $\left[  \delta_{m},\chi^{0}\right]  $ where $dt=105\mu
s$ and (b)
$\left[  \delta_{m},dt\right]  $, where $\chi^{0}=6$. Here, $\Omega_{d}%
^{0}=10^{7}$ s$^{-1}$, $t_{p}=75$ $\mu s$. Lighter areas correspond
to high efficiencies. Pure white areas denote efficiencies more than
$80\%$. The other
parameters are listed in the text.}%
\label{optimized}%
\end{center}
\end{figure}

Figure \ref{optimized}(a) and (b) show the contour plots of the
final molecular productions for sets of $\left[
\delta_{m},\chi^{0}\right]  $ and $\left[  \delta_{m},dt\right]  $,
respectively, where the lighter areas correspond to higher efficiencies.
Especially, pure white regimes are for $\eta\geq80\%$. These two
mappings have several attractive features. Firstly, a symmetric
pattern along the $\delta_{m}$ direction is explicitly
observable, which
can be ascribed to the existence of a three-photon resonance [equation (\ref{delta_g}%
)]. Regardless of whether $\delta_{m}$ is positive or negative, either
$\delta _{g+}=-2\mu_{+}$ or $\delta_{g-}=-2\mu_{-}$ will be
satisfied. In other words, a double-resonant condition must hold on
both sides of $\delta_{m}$, leading to a symmetric double-peak
pattern. Similar patterns have been demonstrated by the
Autler-Townes splitting effect \cite{Aulter55,Bauer09}, which
usually takes place when an optical field is detuned close to an
exact transition frequency. To be more understandable, if we
artificially add a small perturbation to a resonance, the
double-peak profile will be correspondingly shifted. Since this
shift employs no improvement in the molecule production, we
will leave this point for future interested readers.

Secondly, if we fix $\left\vert \delta_{m}\right\vert $ around
$0.05\times 10^{7}$ s$^{-1}$ and gradually increase the values of
$\chi^{0}$ and $dt$, the final transfer efficiencies express similar
variations. Seen in figure \ref{optimized}(a), based on
$\Omega_{d}^{0}=10^{7}$s$^{-1}$, $dt=105\mu s$, if $\chi^{0}$
changes from 1.0 to 6.0, a dramatic enhancement for $\eta$ is
explicit. When further increasing $\chi^{0}$ up to 10.0, $\eta$ values
will be slowly decreasing. A similar trend with $\eta$ as the pulse
delay $dt$ varies
is depicted in figure \ref{optimized}(b) where $\chi^{0}=6$, $\Omega_{d}%
^{0}=10^{7}$s$^{-1}$. When $\delta_{m}=0$, efficiencies are very
poor, which agrees with our CPT predictions equation (\ref{CPT_re})
in section 2. Finally, we find that the base value of
$\Omega_{d}^{0}$ offers few contributions to the transfer. If
$\Omega_{d}^{0}$ is set as $2\times 10^{7}$s$^{-1}$, we will
obtain a much analogous contour plot to figure
\ref{optimized}(a)(not shown).

A brief conclusion for the sections 4.1 and 4.2 is that we are
provided with rich ways to select relevant parameters for
optimal atom-molecule conversion.

\subsection{Population dynamics}

\begin{figure}
[ptb]
\begin{center}
\includegraphics[
height=4.0271in, width=3.0577in
]%
{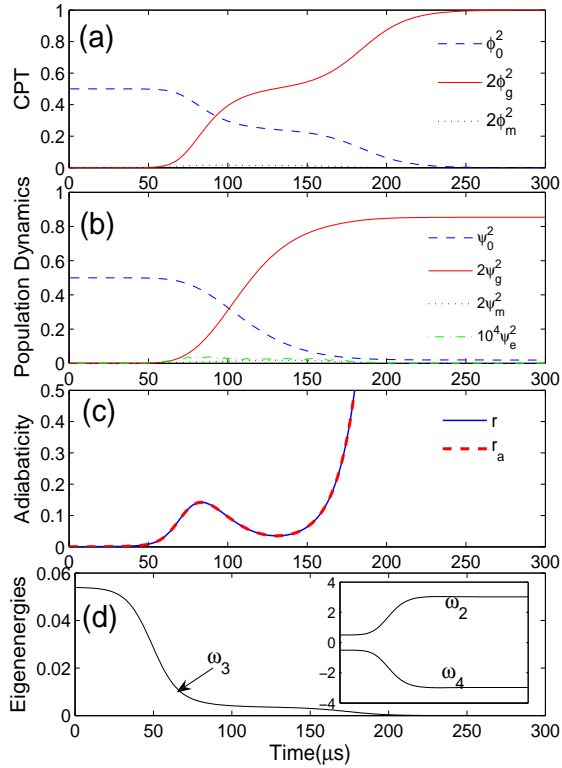}%
\caption{(color online) (a) CPT dynamics, (b) population dynamics,
(c) time-dependent adiabaticity values, (d) three excited
eigen-energies as time changes where $\omega_{3}$ is at least two
orders smaller than $\omega_{2,4}$
(shown in the inset).}%
\label{dy_case}%
\end{center}
\end{figure}

In the following, we consider a concrete example in our "$R$-type"
scheme using the parameters based on our previous discussions.
Optimal parameters are given
by $T=30$ $\mu s$, $\Omega_{PA}^{0}=4\times10^{5}$ s$^{-1}$, $\Omega_{p}%
^{0}=6\times10^{7}$ s$^{-1}$, $\Omega_{d}^{0}=10^{7}$ s$^{-1}$,
$t_{p}=75$ $\mu s$, $t_{d}=180$ $\mu s$, $\left\vert
\delta_{m}\right\vert =5.4\times 10^{5}$ s$^{-1}$,
$\gamma_{m}=3\times10^{4}$ s$^{-1}$ and $\gamma_{e}=10^{7}$
s$^{-1}$. Numerical results are plotted in figure \ref{dy_case}. By
directly integrating the mean-field dynamic equations
(\ref{dy1})-(\ref{dy5}), we produce a population dynamics which
contains all the field amplitudes in figure \ref{dy_case}(b).
Observably, more than 85\% of the atoms $\psi_{0_{1,2}}^{2}$ convert
into ground-state molecules $\psi_{g}^{2}$. Compared with the CPT
dynamics shown in figure \ref{dy_case}(a), a good agreement is
clearly seen, except for a slightly lower $2\psi_{g}^{2}$ coming from
spontaneous decays. In particular, we need to mention that
the $\phi_{m}^{2}(\psi_{m}^{2})$ amplitude (dotted) has been deeply
suppressed, with a maximum value smaller than 0.02.

Figure \ref{dy_case}(c) represents the adiabaticity defined in
equation (\ref{r_value}) as time changes (in solid). By solving
equations (\ref{coupled}) numerically, a complete $r$-value is able
to be determined from the $c_{j}$ values. Three excited eigenenergies
obtained from equation (\ref{cubic}) are displayed in figure
\ref{dy_case}(d) and the inset, where $\omega_{3}$ is smaller than
$\omega_{2,4}$ by orders of magnitude. In the dressed state picture,
$\omega_{j}$ stands for the energy of the $j$th eigenstate, and generally
speaking, a higher-energy eigenstate is usually more difficult to
populate than a lower one. Thereby, in deriving the adiabaticity, we
shall safely neglect the contributions from $\omega_{2,4}$ and
$\mathbf{w}_{2,4}$, simplifying equations
(\ref{coupled}) with $\omega_{3},\mathbf{w}_{3}$ only, yielding%

\begin{equation}
i\omega_{3}c_{3}+f_{33}c_{3}=-\mathbf{\mathbf{w}}_{3}^{T}\mathbf{\dot{\Phi}},
\label{c3}%
\end{equation}
which leads to a reduced assessment on adiabaticity:
$r_{a}=\left\vert
\mathbf{\mathbf{w}}_{3}^{T}\mathbf{\dot{\Phi}/(}i\omega_{3}+f_{33})\right\vert
/\sqrt{3}$. Clearly, $r_{a}$ matches with $r$ in figure \ref{dy_case}(c) perfectly.%

\begin{figure}
[ptb]
\begin{center}
\includegraphics[
height=2.1649in, width=2.8636in
]%
{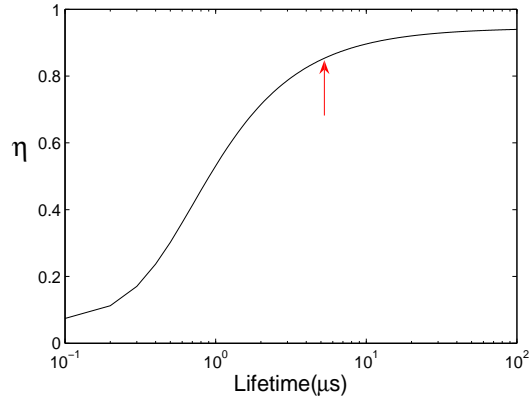}%
\caption{The final transfer efficiency $\eta$ as a function of the
intermediate state lifetime $\tau_{m}$. An arrow corresponds to the lifetime we have used.}%
\label{lifetime}%
\end{center}
\end{figure}

One critical concern in our scheme is the stability of state
$\left\vert m\right\rangle $, which indeed plays a vital role in
determining the final transfer efficiency. In our calculations, we
use the $\left\vert m\right\rangle $ state lifetime to be
$\tau_{m}=1/(2\pi\gamma_{m})=5.3\mu s$, which is comparable with
the earlier work of Napolitano \textit{et al} \cite{Napolitano94}.
Figure
\ref{lifetime} displays how the final efficiency $\eta$(=2$\psi_{g}^{2}%
(\infty)$) varies as a function of $\tau_{m}$. Clearly, if
$\tau_{m}$ is smaller than $1\mu s$, $\eta$ drops rapidly as
$\tau_{m}$ becomes shorter. However, if we are able to find a more
stable intermediate state, with a lifetime longer than 10 $\mu s$,
the corresponding transfer efficiency reaches as high as 90\%. The
arrow shown in figure \ref{lifetime} points to the
$\tau_{m}$ value used in our paper.%

Finally, it is meaningful for a numeric estimate of the feasibility of our
scheme by taking the KRb molecule as a possible candidate in experiment. Based
on the predictions in \cite{Aikawa09,Beuc06}, it is experimentally possible
to find out an appropriate $\left\vert m\right\rangle $ state with a
relative long lifetime, e.g. $3^{1}\Sigma ^{+}$. There is a transition
dipole moment of $\sim $ 1 ea$_{0}$ for the state $3^{1}\Sigma ^{+}$, which
corresponds to the lifetime of several $\mu s$. In addition, the $1^{1}\Pi $
is a good candidate for the high-lying $\left\vert e\right\rangle $ state
because of its purely singlet character and favorable transition dipole
moment associated with the lowest singlet state $1^{1}\Sigma ^{+}$ ($%
\left\vert g\right\rangle $ state). A rough estimation of the PA power,
adopting the parameters of the free-bound FCF $I_{FCF}\sim $10$^{-14}$ m$%
^{3/2}$ for KRb \cite{Drummond02}-\cite{Dulieu04} and an initial atomic
density $n_{0}=10^{20}$m$^{-3}$, gives rise to a PA laser intensity of $%
I_{PA}=2c\varepsilon _{0}(\hbar \Omega _{PA}^{0(el)}/\mu )^{2}\approx 512.7$
W/cm$^{2}$ for our "R-type" scheme, where $c$ is the light velocity, $%
\varepsilon _{0}$ is vacuum permittivity, $\mu $ is the dipole moment, $%
\Omega _{PA}^{0(el)}$ is the electronic Rabi frequency defined by $\Omega
_{PA}^{0(el)}=\Omega _{PA}^{(0)}/(I_{FCF}\sqrt{n_{0}})$ \cite{Drummond02}.
However, in a standard "$\Lambda $-type" system (see figure \ref{model}),
the absence of state $\left\vert m\right\rangle $ requires a more intense PA
field to stimulate particles into the highly excited state $\left\vert
e\right\rangle $. As a result, to achieve the same production rate of
molecules as in our "R-type" case, the required PA power must be $\Omega
_{PA}^{\prime 0}\sim 6.4\times 10^{6}$ s$^{-1}$. The corresponding PA
intensity is  $I_{PA}^{\prime }\sim 1.31\times 10^{5}$ W/cm$^{2}$, giving
other parameters the same as in the "R-type" case. Evidently, the above
numeric estimate shows a more than 250 times power reduction in our "R-type"
approach.

\section{Summary}

Although the magnetic FR-assisted PA technique has been considered as the
most promising way for the purpose of overcoming the PA weakness, the
primary drawback in such a scheme is the strong inelastic-collisional loss
of Feshbach molecules, especially when the magnetic field is tuned near the
resonant point. To eliminate the bottleneck in the magnetic FR-assisted PA
technique, we work out a robust all-optical atom-molecule conversion model
through a "$R$-type" photoassociative STIRAP, where an intermediate state $%
\left\vert m\right\rangle $ is introduced to form a quasi-CPT state. In
terms of the detailed adiabatic theorem, we show that the quasi-CPT state
can lead to a higher atom-molecule transfer efficiency with a lower PA laser
power, compared to the normal CPT state in the conventional all-optical "$%
\Lambda $-type" two-color PA configuration. 

The key reason for the lowered
power of PA laser is due to the existence of an intermediate state $%
\left\vert m\right\rangle $. In this case, it is easier to photoassociate
free atoms into this low-lying $\left\vert m\right\rangle $ state, rather
than a high-lying $\left\vert e\right\rangle $ state. Since molecules in
state $\left\vert m\right\rangle $ are unstable, the subsequent STIRAP
transfer from the intermediate state $\left\vert m\right\rangle $ to the
final state $\left\vert g\right\rangle $ must be rapid enough to avoid the
loss of molecules from the $\left\vert m\right\rangle $ state, here
characterized by a finite CPT lifetime. In addition, we also show that the
lifetime of state $\left\vert m\right\rangle $ will significantly affect the
final transfer efficiency. A specific estimation to illustrate the
feasibility of our approach is performed. Finally we want to emphasize that
the scheme proposed here is the first one to overcome the inefficiency of PA
with only all-optical fields involved. This may open up new opportunities
for experimental endeavors to create polar molecular condensates directly
from ultracold atoms. A more careful treatment taking into account nonlinear
collisions will be left for future explorations.

\section*{Acknowledgments}

This work is supported by the National Natural Science Foundation of China
under Grant No. 10588402, the National Basic Research Program of China (973
Program) under Grant No. 2006CB921104, the Program of Shanghai Subject Chief
Scientist under Grant No. 08XD14017, and the Shanghai Leading Academic
Discipline Project under Grant No. B480 (W.Z.), the National Natural Science
Foundation of China under Grant No. 10974057 and No. 10874045, Shanghai
Pujiang Program under Grant No. 08PJ1405000 (L.Z.).

\appendix
\section{$\mathbf{Q}$ vector}
$q_{i}$ values in vector $\mathbf{Q}$ are given by%

\begin{eqnarray}
q_{0} &  =-\frac{\phi_{0}^{2}+2\phi_{m}^{2}\left(  1+\chi^{2}\right)
}{2v\Omega_{PA}\phi_{0}\phi_{m}^{2}}\label{q0}\\
q_{m} &  =-\frac{1}{v\Omega_{PA}\phi_{m}}\label{qm}\\
q_{e} &  =\frac{2\chi}{v\Omega_{d}}\label{qe}\\
q_{g} &  =\frac{\chi}{v\Omega_{PA}\phi_{m}}\label{qg}%
\end{eqnarray}
and

\begin{equation}
v=-\frac{\phi_{0}^{2}+4\phi_{m}^{2}(1+\chi^{2})}{\Omega_{PA}\phi_{m}^{3}}\label{v}%
\end{equation}


\section*{References}
\begin{thebibliography}{10}

\bibitem {Doyle04} Doyle J, Friedrich B, Krems R V and Masnou-Seeuws F, 2004
\textit{Eur. Phys. J. D} \textbf{31}, 149

\bibitem {Summary}Carr L D, DeMille D, Krems R V and Ye J 2009 \textit{New J.
Phys.} \textbf{11}, 055049

\bibitem {Santos00}Santos L, Shlyapnikov G V, Zoller P, and Lewenstein M 2000
\textit{Phys. Rev. Lett.} \textbf{85}, 1791

\bibitem {Syi00}Yi S and You L 2000 \textit{Phys. Rev. A.} \textbf{61}, 041604(R)

\bibitem {DeMille02}DeMille D 2002 \textit{Phys. Rev. Lett.} \textbf{88}, 067901

\bibitem {Zoller06}Andr\'{e} A \textit{et al. }2006 \textit{Nat. Phys.}
\textbf{2}, 636

\bibitem {Yelin06}Yelin S F, Kirby K and C\^{o}t\'{e} R 2006 \textit{Phys. Rev.
A.} \textbf{74}, 050301(R)

\bibitem {Kozlov95}Kozlov M G and Labzowsky L N 1995 \textit{J. Phy. B},
\textbf{28}, 1933

\bibitem {DeMille00}DeMille D \textit{et al.} 2000 \textit{Phys. Rev. A.}
\textbf{61}, 052507

\bibitem {Hudson02}Hudson J J, Sauer B E, Tarbutt M R, and Hinds E A 2002
\textit{Phys. Rev. Lett.} \textbf{89}, 023003

\bibitem {Hudson06}Hudson E R, Lewandowski H J, Sawyer B C, and Ye J, 2006
\textit{Phys. Rev. Lett.} \textbf{96}, 143004

\bibitem {Flambaum07}Flambaum V V and Kozlov M G 2007 \textit{Phys. Rev.
Lett.} \textbf{99}, 150801

\bibitem {DeMille08}DeMille D \textit{et al.} 2008 \textit{Phys. Rev. Lett.}
\textbf{100}, 023003

\bibitem {Doyle98}Weinstein J D \textit{et al. }1998 \textit{Nature}
\textbf{395}, 148

\bibitem {Jones06}Jones K M, Tiesinga E, Lett P D and Julienne P S 2006
\textit{Rev. Mod. Phys}, \textbf{78}, 483

\bibitem {Thorsten06}K\"{o}hler T, G\'{o}ral K and Julienne P S 2006
\textit{Rev. Mod. Phys}, \textbf{78}, 1311

\bibitem {Bergmann98}Bergmann K, Theuer H, and Shore B W, 1998 \textit{Rev.
Mod. Phys}, \textbf{70}, 1003

\bibitem {Winkler07}Winkler K \textit{et al.} 2007 \textit{Phys. Rev. Lett.}
\textbf{98}, 043201

\bibitem {Lang08}Lang F \textit{et al.} 2008 \textit{Phys. Rev. Lett.}
\textbf{101}, 133005

\bibitem {Ospelkaus08}Ospelkaus S \textit{et al.} 2008 \textit{Nat. Phys.
}\textbf{4} 622

\bibitem {KKNi08}Ni K-K \textit{et al.} 2008 \textit{Science}, \textbf{322}, 231

\bibitem {Ospelkaus0811}Ospelkaus S \textit{et al.} 2009 \textit{Faraday Discuss.}, \textbf{142}, 351

\bibitem {Danzl08}Danzl J G \textit{et al.} 2008 \textit{Science},
\textbf{321}, 1062

\bibitem {Aikawa09}Aikawa K, Akamatsu D, Kobayashi J, Ueda M, Kishimoto T and
Inouye S 2009 \textit{New J. Phys.} \textbf{11}, 055035

\bibitem {Wynar00}Wynar R, Freeland R S, Han D J, Ryu C and Heinzen D J 2000
\textit{Science}, \textbf{287}, 1016

\bibitem {Tolra01}Tolra B L, Drag C, and Pillet P 2001 \textit{Phys. Rev. A.}
\textbf{64}, 061401(R)

\bibitem {Sage05}Sage J M, Sainis S, Bergeman T, and DeMille D 2005
\textit{Phys. Rev. Lett.} \textbf{94}, 203001

\bibitem {Kerman04}Kerman A J, Sage J M, Sainis S, Bergeman T, and DeMille D
2004 \textit{Phys. Rev. Lett.} \textbf{92}, 033004

\bibitem {Pillet08}Viteau M 2008 \textit{Science}, \textbf{321}, 232

\bibitem {Deiglmayr08}Deiglmayr J \textit{et al.} 2008 \textit{Phys. Rev.
Lett.} \textbf{101}, 133004

\bibitem {Vardi97}Vardi A, Abrashkevich D, Frishman E, and Shapiro M 1997
\textit{J. Chem. Phys.} \textbf{107}, 6166

\bibitem {Vardi99}Vardi A, Shapiro M and Bergmann K, 1998 \textit{Opt.
Express.} \textbf{4}, 91

\bibitem {Abeelen98}van Abeelen F A, Heinzen D J and Verhaar B J 1998
\textit{Phys. Rev. A.} \textbf{57}, r4102

\bibitem {Courteille98}Courteille Ph, Freeland R S, Heinzen D J, van Abeelen F
A and Verhaar B J 1998 \textit{Phys. Rev. Lett.} \textbf{81}, 69

\bibitem {Tolra03}Tolra B L \textit{et al.} 2003 \textit{Europhy. Lett.}
\textbf{64}, 171

\bibitem {Junker08}Junker M \textit{et al.} 2008 \textit{Phys. Rev. Lett.}
\textbf{101}, 060406

\bibitem {Deiglmayr09}Deiglmayr J \textit{et al.} 2009 \textit{New J. Phys},
\textbf{11}, 055034

\bibitem {Pellegrini08}Pellegrini P, Gacesa M, and C\^{o}t\'{e} R 2008
\textit{Phys. Rev. Lett.} \textbf{101}, 053201

\bibitem {Kuznetsova09}Kuznetsova E, Gacesa M, Pellegrini P, Yelin S F and
C\^{o}t\'{e} R 2009 \textit{New J. Phys}, \textbf{11}, 055028

\bibitem {Nikolov00}Nikolov A N, Ensher J R, Eyler E E, Wang H, Stwalley W C
and Gould P L 2000 \textit{Phys. Rev. Lett.} \textbf{84}, 246

\bibitem {Band95}Band Y B and Julienne P S 1995 \textit{Phys. Rev. A.}
\textbf{51}, R4317

\bibitem {Kokkelmans01}Kokkelmans S J J M F, Vissers H M J and Verhaar B J
2001 \textit{Phys. Rev. A.} \textbf{63}, 031601

\bibitem {Mackie02}Mackie M 2002 \textit{Phys. Rev. A.} \textbf{66}, 043613

\bibitem {Heinzen00}Heinzen D J, Wynar R, Drummond P D and Kheruntsyan K V
2000 \textit{Phys. Rev. Lett.} \textbf{84}, 5029

\bibitem {Napolitano94}Napolitano R, Weiner J, Williams C. J. and Julienne P.
S. 1994 \textit{Phys. Rev. Lett.} \textbf{73}, 1352

\bibitem {Kuznetsova08}Kuznetsova E, Pellegrini P, C\^{o}t\'{e} R, Lukin M D
and Yelin S F 2008 \textit{Phys. Rev. A.} \textbf{78}, 021402(R)

\bibitem {Pu07}Pu H, Maenner P, Zhang W P and Ling H Y 2007 \textit{Phys. Rev.
Lett.} \textbf{98}, 050406

\bibitem {Jing08}Jing H, Zheng F, Jiang Y, and Geng Z 2008 \textit{Phys. Rev.
A.} \textbf{78}, 033617

\bibitem {Koenig04}L-Koenig E, Kosloff R, Masnou-Seeuws F, and Vatasescu M 2004
\textit{Phys. Rev. A.} \textbf{70}, 033414

\bibitem {Ling07}Ling H Y, Maenner P, Zhang W P, and Pu H 2007 \textit{Phys.
Rev. A.} \textbf{75}, 033615

\bibitem {Aulter55}Aulter S. H and Townes C. H 1955 \textit{Phys. Rev.}
\textbf{100}, 703

\bibitem {Bauer09}Bauer D M, Lettner M, Vo C, Rempe G and D\"{u}rr S 2009
\textit{Nature Phys.} \textbf{5}, 339 ; \textit{ibid}. 2009
\textit{Phys. Rev. A}. \textbf{79}, 062713

\bibitem {Beuc06}Beuc R, Movre M, Ban T, Pichler G, Aymar M, Dulieu O and Ernst W E 
2006 \textit{J. Phys. B: At. Mol. Opt. Phys.} \textbf{39}, S1191

\bibitem {Drummond02}Drummond P D, Kheruntsyan K V, Heinzen D J and Wynar R H
2002 \textit{Phys. Rev. A.} \textbf{65}, 063619

\bibitem {Naidon03}Naidon P and Masnou-Seeuws F 2003 \textit{Phys. Rev. A.}
\textbf{68}, 033612

\bibitem {Dulieu04}Azizi S, Aymar M and Dulieu O 2004 \textit{Eur. Phys. J. D}
\textbf{31}, 195
\end{thebibliography}
\end{document}